\documentstyle[12pt]{article}
\newcommand{\rf}[1]{(\ref{#1})}
\newcommand{\beq}{\begin{equation}}
\newcommand{\eeq}{\end{equation}}
\newcommand{\bea}{\begin{eqnarray}}
\newcommand{\eea}{\end{eqnarray}}
\newcommand{\x}{{\bf x}}

\newcommand{\k}{\kappa}
\newcommand{\e}{\epsilon}
\newcommand{\et}{{\tilde e}}
\newcommand{\wt}{{\tilde \omega}}
\newcommand{\wb}{{\overline \omega}}

\newcommand{\p}{\phi}
\newcommand{\w}{\omega}

\renewcommand{\d}{\delta}
\renewcommand{\l}{\lambda}
\renewcommand{\b}{\beta}
\renewcommand{\a}{\alpha}
\newcommand{\n}{\nu}
\newcommand{\m}{\mu}
\newcommand{\r}{\rho}
\newcommand{\s}{\sigma}

\newcommand{\oh}{\frac{1}{2}}
\newcommand{\oq}{\frac{1}{4}}

\newcommand{\non}{\nonumber}

\newcommand{\zp}{\frac{d^4p}{(2\pi)^4}}
\newcommand{\ra}{\rightarrow}
\renewcommand{\t}{\tau}
\begin{document}
\topmargin 0pt
\oddsidemargin 5mm
\headheight 0pt
\headsep 0pt
\topskip 9mm
\addtolength{\baselineskip}{0.20\baselineskip}
\hfill SFSU-TH-92/1

\hfill April 1992
\bigskip
\bigskip

\begin{center}

\vspace{36pt}
{\large \bf STABILIZED QUANTUM GRAVITY:}
\end{center}
\begin{center}
{\large \bf Stochastic Interpretation and Numerical Simulation}
\footnote{Supported by the U.S. Department of Energy under
Contract No. DE-AC03-81ER40009.}
\end{center}

\vspace{36pt}

\begin{center}
{\sl J. Greensite}

\vspace{12pt}

Physics and Astronomy Dept.\\
San Francisco State University\\
1600 Holloway Ave.\\
San Francisco, CA 94132\\

\end{center}
\vfill

\begin{center}
{\bf Abstract}
\end{center}

\vspace{12pt}

     Following the reasoning of Claudson and Halpern, it is shown
that "fifth-time" stabilized quantum gravity is equivalent to Langevin
evolution (i.e. stochastic quantization) between fixed non-singular,
but otherwise arbitrary, initial and final states.  The simple
restriction to a fixed final state at $t_5 \rightarrow \infty$ is
sufficient to stabilize the theory.  This equivalence fixes the
integration measure, and suggests a particular operator-ordering, for
the fifth-time action of quantum gravity.  Results of a numerical
simulation of stabilized, latticized Einstein-Cartan theory on some
small lattices are reported.  In the range of cosmological constant $\l$
investigated, it is found that: 1) the system is always in the broken
phase $<det(e)> \ne 0$; and 2) the negative free energy is large, possibly
singular, in the vincinity of $\l = 0$.  The second finding may be
relevant to the cosmological constant problem.

\vspace{24pt}

\vfill

\newpage

\section{Introduction}

   The "fifth-time" action is a general procedure, first proposed in
ref. \cite{GH}, for stabilizing theories whose Euclidean actions
are unbounded from below.  There are a several such "bottomless action"
theories which are of interest to physics; examples include D=0 matrix
models, bosonic string field theory, and, especially, Einstein-Hilbert
gravity.  The fifth-time method has been applied by many authors
\cite{matmod} to D=0 matrix models.  In this article I will continue
the study of fifth-time stabilized quantum gravity, begun in ref.
\cite{Me1}.

   Euclidean quantum gravity is of interest for at least two reasons.
First, there are the intriguing Baum-Hawking-Coleman arguments \cite{BHC}
for the vanishing of the cosmological constant, which are formulated
in the Euclidean approach.  Unfortunately, these arguments involve the
semiclassical evaluation of a badly divergent path-integral, and are
therefore rather suspect \cite{Unruh}.  Secondly, Monte Carlo simulations
of quantum gravity cannot avoid the Euclidean formulation.
In ref. \cite{Me1} it was shown how the fifth-time prescription
generates a new diffeomorphism-invariant action
for quantum gravity, bounded from below,
which leads to the usual Einstein equations of
motion in the classical limit.  Moreover, the stabilized Einstein-Cartan
theory appears to be reflection-positive, at least in lattice formulation,
and should be a good starting point for non-perturbative investigations.

  At the perturbative level, the stabilized theory flips the sign of
the "wrong-sign" conformal mode
at zeroth order, and is non-local at higher orders.  This behavior
is in agreement with two other approaches which aim at deriving the
Euclidean theory from Minkowski space gravity. The
first, due to Schleich \cite{Schleich} (see also \cite{Kosower}),
solves the Hamiltonian constraints in the Minkowski theory prior to
Wick rotation, and then reinserts redundant degrees of freedom to
recover diffeomorphism invariance.  The second approach, due to
Biran et. al. \cite{Brout} and Mazur and Mottola \cite{Mazur}, argues
that the kinetic term of the
conformal factor should be Wick rotated like a potential term, when
certain Jacobian factors in the integration measure are taken into
account.  In both cases a bounded, non-local Euclidean action is generated.
These studies show that the Euclidean continuation of Einstein-Hilbert
gravity is not necessarily obtained by the naive replacement of the Minkowski
action by the corresponding (unbounded) Euclidean expression.  The
advantage of the fifth-time approach is that the 5-dimensional action
is local, relatively simple, and amenable to numerical and analytical
methods, whereas the non-local stabilized 4-dimensional action, obtained
in any of these approaches, is not known in closed form.

   A common objection to the fifth-time prescription is that it is
rather ad hoc, at least as presented in the original treatment in
\cite{GH}.  The method generates a stabilized action which has the
same classical limit and formal perturbative expansion as the
unstabilized theory, but there could be other methods to achieve
those ends.  In fact, for D=0 matrix models alternate stabilizations
have been proposed \cite{etal,Spaniards}.
A better understanding of the physics
underlying the fifth-time action would therefore be helpful in judging
its merits.  In the next section it will be shown that the fifth-time
prescription is equivalent to stochastic quantization, with the
constraint that the Langevin evolution not only begin at a fixed
(arbitrary) initial state (at $t_5 \rightarrow - \infty$),
but also {\it terminate} at an arbitrary non-singular final state
($t_5 \ra \infty$).  No claim of originality is made for this
"stochastic interpretation" of the fifth-time action.  The
equivalence of Langevin evolution to the fifth-time action, for
ordinary bounded actions, was shown by Gozzi in ref. \cite{Gozzi}.
The idea of fixing the final state of Langevin evolution for
bottomless actions is due to Claudson and Halpern \cite{HC}, in
connection with Nicolai maps for $QCD_4$, and it was noted by
Giveon et. al. \cite{Giveon} that Langevin evolution with fixed
initial and final states is equivalent to the fifth-time stabilization
prescription of ref. \cite{GH}.  Section 2 contains an exposition
of this reasoning, which  will then be
extended, in section 3, to the special case of quantum gravity. It will
be shown that the stochastic interpretation fixes the integration
measure, and suggests a particular choice of operator-ordering,
in the fifth-time action.

   In section 4 the results of a Monte Carlo simulation of stabilized,
latticized, Einstein-Cartan gravity are reported.  The latticization is
similar to that proposed by Menotti and Pelissetto in ref. \cite{Menotti},
in that local O(4) symmetry is exact, but diffeomorphism invariance
is broken by the lattice regularization.  The simulation is carried out
on very tiny lattices: $2^4 \times 4$ and $3^4 \times 4$; so obviously
the results must viewed with caution.  Nevertheless, the outcome is
interesting.  It is found, for a range of positive and negative
cosmological constants, that the system is always in the broken phase
$<det(e)> \ne 0$.  Moreover, in the neighborhood of $\l = 0$: the
average curvature approaches 0, the volume per lattice site diverges,
and there is a peak, possibly divergent, in the
negative free energy/lattice site $-F$.
The peak in $-F$ near $\l=0$ and infinite volume suggests that,
if the cosmological constant becomes dynamical by some mechanism,
then its probability distribution is peaked near $\l=0$, as proposed in
ref. \cite{BHC}. Recently Carlini and Martellini \cite{Martellini}
have argued that the Coleman mechanism for a vanishing cosmological term
is realized in fifth-time stabilized quantum gravity; this could be
related to the numerical results found here.

  An alternative approach to formulating and simulating Euclidean
quantum gravity is the method of summing over simplicial manifolds.
In this case the action is just the Euclidean Einstein-Hilbert
(Regge) action, and it is the lattice regularization (restriction to
simplicial manifolds) which provides a cutoff on the
number of manifolds with large positive or negative curvature
\cite{Migdal}.  This lattice cutoff on large curvatures
makes numerical simulation possible, and
interesting Monte Carlo results have recently been obtained
in 4-dimensions [17-20].
It remains to be seen whether a theory whose stability is based
on the lattice structure is actually independent of the lattice
structure, in the sense of universality. Also, as noted above, the
continuation of Minkowski space
gravity to Euclidean space is not necessarily obtained by replacing
the Minkowski space Einstein-Hilbert action by the Euclidean
Einstein-Hilbert action.  Some additional remarks are contained in
section 4.

\section{Stochastic Quantization of Bottomless Actions}

   This section reviews a selection of ideas found in ref. [12-14],
which provide a stochastic interpretation for the fifth-time
action prescription of ref. [1].

   What does it mean to quantize a bottomless action $S[\phi]$ in
Euclidean space?  Clearly, if the Boltzman factor $exp(-S)$ is
non-integrable, the path-integral formulation cannot be used
directly.  However, in statistical physics the Boltzman factor
is just a representation of the probability distribution obtained
by random Brownian motion.  From this point of view it is the
Brownian motion, as described by the Langevin equation, which is
fundamental.  Quantization based on the Langevin equation, known
as "stochastic quantization", gives the following prescription
for thermal averages of operators $Q[\p]$:

\beq
    <Q> = \lim_{T \ra \infty} {1 \over Z_5}
\int D\eta(\x,-T<t_5<T) \; Q[\phi(\x,0)]
              exp[-\int^T_{-T} d^5x \; \eta^2/4 \hbar]
\label{I}
\eeq
where
\bea
     \partial_5 \phi &=& -{\d S \over \d \p} + \eta
\non \\
     \phi(\x,-T) &=& \p_i(\x)
\eea
In stochastic quantization, the role of the classical action $S$
is to supply the drift term in the Langevin equation.
For a given noise configuration $\eta$, the Langevin equation is
solved starting from an initial field configuration
$\p_i(\x)$ at $t_5=-T$. As $T \ra \infty$, the thermal average should be
independent of the initial configuration.

  Applied to the problem of bottomless actions, the Langevin approach
does not, at first sight, seem much of an improvement over path-integral
quantization.  Even if the initial configuration $\p_i$ is a local
minimum of the action, for large $t_5$ the field will almost always
be driven into the bottomless region of the action, evolving
towards a singular configuration as $t_5 \ra \infty$.  The qualifier
in "almost always" is important, however, since there do exist
noise configurations $\eta(\x,t_5)$ for which the field remains in
the vicinity of a stationary point $\d S / \d \p = 0$ throughout
its Langevin evolution.

   The proposal, then, is as follows:  Let us retain the prescription of
stochastic quantization that $S[\p]$, bottomless or not, supplies the
drift term of Langevin evolution, but average only over noise terms which
leave the field in a fixed non-singular final state $\p_f(\x)$ at
$t_5=T$.  In other words,

\bea
    <Q> &=& \lim_{T \ra \infty} {1 \over Z_5}
\int  D\eta(x,-T<t_5<T)  \; \d[\p(\x,T)-\p_f(\x)] Q[\phi(\x,0)]
\non \\
   & & \times     e[-\int^T_{-T} d^5x \; \eta^2/4 \hbar]
\label{IF}
\eea
where, again,
\bea
     \partial_5 \phi &=& -{\d S \over \d \p} + \eta
\non \\
     \phi(\x,-T) &=& \p_i(\x)
\eea
If we can show that thermal averages $<Q>$ are
independent of the choice of both initial $\p_i$ {\it and} final $\p_f$
configurations, as $T \ra \infty$,  then we have obtained a version
of stochastic quantization which, for bounded actions, is equivalent
to the usual version, but which is also meaningful for bottomless actions.
I will refer to the standard version of eq. \rf{I}, with a fixed
initial state, as "stochastic (I)" quantization, and the proposal of eq.
\rf{IF}, with fixed initial and final states, as "stochastic (IF)"
quantization.

   Following ref. \cite{Gozzi,HC}, let us change variables
in the functional integral of eq. \rf{IF}.  Writing the Langevin
equation as

\beq
      \eta = \partial_5 \p + {\d S \over \d \p}
\eeq
we have

\bea
      Z_5 &=& \int D\eta \; \d[\p(\x,T)-\p_f(\x)]
exp[-\int^T_{-T} d^5x \; \eta^2/4\hbar]
\non \\
        &=& \int^{\p_f}_{\p_i} D\p \; det[{\d \eta \over \d \p}]
exp\left[-\int_{-T}^T d^5x \; [(\partial_5 \p)^2 + ({\d S \over \d \p})^2
+ 2 \partial_5 \p {\d S \over \d \p}]/4\hbar \right]
\non \\
        &=& \int^{\p_f}_{\p_i} D\p \; det[{\d \eta \over \d \p}]
exp\left[-\int_{-T}^{T} d^5x \;[\oq (\partial_5 \p)^2 +
\oq ({\d S \over \d \p})^2]/\hbar \right]
\non \\
  & & \times e^{-(S[\p_f]-S[\p_i])/2\hbar}
\label{etaint}
\eea

Working out the Jacobian,

\bea
   \lefteqn{ det[{\d \eta(x) \over \d \p(x')}] }
\non \\
&=& det[\partial_5 \d^5(x-x') + {\d^2S \over \d\p(\x) \d\p(\x')}_{|t_5}
\delta(t_5-t'_5)]
\non \\
&=& det[\partial_5(\d^5(x-x') + \theta(t_5-t'_5){\d^2S \over \d\p(\x)
\d\p(\x')}_{|t_5})]
\non \\
&=& det[\partial_5] det[\d^5(x-x') + \theta(t_5-t'_5) {\d^2S \over
\d\p(\x) \d\p(\x')}_{|t_5}]
\non \\
&=& det[\partial_5] exp\left[ Trln[\d^5(x-x') + \theta(t_5-t'_5)
{\d^2 S \over \d\p(\x) \d\p(\x')}_{|t_5}] \right]
\non \\
&=& det[\partial_5] exp\left[ \oh \int d^5x \; {\d^2 S \over \d\p^2} \right]
\label{det1}
\eea
where $\theta(0)=\oh$, and the meaning of

\beq
     {\d^2 S \over \d\p(\x) \d\p(\x')}_{|t_5}
\eeq
is to carry out the D=4 dimensional functional variations, and then
replace $\p(\x)$ by $\p(\x,t_5)$.  Note that only the first term
survives in the expansion of the Trace log; all other terms in the trace
vanish due to the time-ordering enforced $\theta$.
Substitution of \rf{det1} and \rf{etaint} into \rf{IF} gives

\bea
     <Q> &=& \lim_{T \ra \infty} {1 \over Z_5}\int^{\p_f}_{\p_i}
D\p(\x,-T<t_5<T) \; Q[\p(\x,0)]
\non \\
& & \times exp[-\int^{T}_{-T} d^5x \; [\oq (\partial_5 \p)^2 + \oq
({\d S \over \d \p})^2 - {\hbar \over 2}{\d^2 S \over\d \p^2}]/\hbar]
\eea
or, with a rescaling of $t_5 \ra t_5/2\hbar$

\bea
       <Q>  &=& {1 \over Z_5} \int D\p \; Q[\p(\x,0)] e^{-S_5}
\non \\
        S_5 &=& \int d^5x \; [ \oh(\partial_5 \p)^2 + {1\over 8 \hbar^2}
({\d S \over \d\p})^2 - {1 \over 4\hbar}{\d^2 S \over \d\p^2}]
\label{S5}
\eea

   It is now easy to prove that $<Q>$ is independent of the initial and
final states $\p_i$ and $\p_f$, since

\bea
      <Q> &=& \lim_{T \ra \infty} {\sum_{nm} \Psi_n[\p_f] <\Psi_n|Q|\Psi_m>
 \Psi^*_m[\p_i] e^{-(E_n+E_m)T/\hbar}
\over
      {\sum_{n} \Psi_n[\p_f]
 \Psi^*_n[\p_i] e^{-2E_n T/\hbar} } }
\non \\
         &=& <\Psi_0|Q|\Psi_0>
\label{Q}
\eea
where the $E_n$ are the eigenvalues, $\Psi_n$ the eigenstates
($\Psi_0$ the ground state),
of the Hamiltonian corresponding to $S_5$

\beq
       H_5 = \int d^4x [ -\oh {\d^2 \over \d\p^2}
+ {1 \over 8 \hbar^2}({\d S \over \d\p})^2 - {1\over 4\hbar}
{\d^2S \over \d\p^2} ]
\eeq
which is known as the Fokker-Planck Hamiltonian.
Note that whether $S$ is bottomless or not, the potential term
in $H_5$ is bounded from below, and therefore has a well-defined
ground state.   Since $<Q>$ depends only on the ground state of
$H_5$, and not on the initial or final configurations $\p_i$,
$\p_f$, we have shown that stochastic (IF) quantization, like ordinary
stochastic (I) quantization, does not depend on the the choice of
the initial/final configurations.

   There is one way that the reasoning above could have failed.
This would be if $\Psi_0[\p_f]=0$,  e.g. if $\p_f$ or its derivatives
were infinite.  Such singular configurations would be obtained
in Langevin evolution for bottomless actions, as $t_5 \ra \infty$,
if the final state were unconstrained.  Thus, the only slight restriction
we need to make on the final configuration is that it is non-singular,
$\Psi_0[\p_f] \ne 0$; in fact, this restriction is implicit in the
choice of initial configuration as well, even for ordinary stochastic (I)
quantization.

    Equation \rf{S5} is the "fifth-time" stabilization prescription proposed in
\cite{GH}, which is now seen to be equivalent to stochastic (IF)
quantization.   It is easy to see from \rf{S5} that the $\hbar \ra 0$ limit
enforces the classical equations of motion $\d S / \d \p = 0$, and it
was shown in \cite{GH} that the perturbative expansion of \rf{S5}
reproduces the naive perturbation expansion, to all orders in any
expansion parameter, generated from Taylor expanding $e^{-S}$ in
the usual functional integral.\footnote{The equality of perturbative
expansions holds providing  $S$ is stable at zeroth order, which is not
the case for quantum gravity, c.f. \cite{Me1}.}

     Writing the ground state $\Psi_0$
of $H_5$ in the form

\beq
     \Psi_0 = exp[-S_{eff}/2\hbar]
\eeq
fifth-time
stabilization can be regarded as replaced the bottomless action
$S$ by a bounded action $S_{eff}$, i.e

\bea
   <Q> &=& <\Psi_0|Q|\Psi_0>
\non \\
       &=& {1 \over Z} \int D\p(\x) Q[\p(\x)] e^{-S_{eff}/\hbar}
\eea
which has the same classical
equations of motion, and the same formal perturbative expansion,
as $S$.

\bigskip

\noindent \underline{\bf The Yang-Mills Vacuum as Stabilized
Chern-Simons Theory}

\bigskip

     A curious example of the 5-th time approach is provided by
Yang-Mills theory.  The discussion in this subsection is based
on the Nicolai map for $QCD_4$ found in ref. [13], but with emphasis laid
on the fact that $QCD_4$ itself can be regarded as a "fourth"-time action.

   It has long been known \cite{Me2} that there is an exact
solution of the Yang-Mills Schrodinger equation in temporal gauge

\beq
       \int d^3x \oh[-{\d^2 \over \d A_i^{a2}} + \oh F_{ij}^{a2}]
\Psi = E \Psi
\label{YMSeq}
\eeq
with energy $E=0$; this is the Chern-Simons state

\bea
        \Psi[A] &=& exp\left[\oh \int d^3x
\e_{ijk}[A^a_i \partial_j A^a_k - {2g\over 3}\e_{abc} A^a_iA^b_jA^c_k] \right]
\non \\
                &=& exp\{cs[A]/2 \}
\eea
But because this state is non-normalizable (the Chern-Simons action
is bottomless), it must be rejected as a physical state.
In fact, due to  asymptotic freedom, the {\it true}
Yang-Mills vacuum can be expected to look something like the
abelian ground state for high frequency fluctuations

\beq
          \Psi_0^{abelian}[A] = exp\left[-{1\over 8\pi^2}
\int d^3x d^3y {F_{ij}(x) F_{ij}(y) \over |x-y|^2} \right]
\eeq
while for low-frequency fluctuations, it was argued in ref.
\cite{Me2} (see also \cite{Feynmann}) that the true Yang-Mills
vacuum has the form

\beq
          \Psi_0[A_{low}] \approx exp[-\m \int d^3x \; TrF_{ij}^2]
\eeq
This expression for the low-frequency vacuum has since been
verified by Monte-Carlo simulations \cite{Me-Arisue}.

   Now although the Chern-Simons state is non-normalizable, it is still
an exact solution of the Yang-Mills Schrodinger equation, and an
interesting question to ask is what does the stabilized theory looks like.
The answer is quite remarkable: {\it Stabilized Chern-Simons theory is the
true Yang-Mills vacuum}!  This is easy to see, since the rule
is to replace the unbounded distribution $exp\{cs[A]\}$ by
$exp[-S_{eff}]$,
where $\Psi_0=exp[-S_{eff}/2]$ is the ground state of the Fokker-Planck
Hamiltonian

\beq
     H_5  = \int d^3x \left[ -\oh {\d^2 \over \d A_i^{a2}}
+ {1\over 8}({\d cs[A] \over \d A^a_i})^2 - \oq {\d^2 cs[A] \over
\d A^{a2}_i} \right]
\eeq
which in this case turns out to be the Hamiltonian of D=4 Yang-Mills theory
in temporal gauge, seen in eq. \rf{YMSeq}. As a  consequence, stabilized
Chern-Simons theory is simply the Yang-Mills vacuum in temporal gauge.
Likewise, the "fifth-time" action (really
a "4-th time" action, since $cs[A]$ is 3-dimensional) is just the D=4
Yang-Mills theory.  From this fact, it is not hard to see
that {\it all} (not just equal-time)
correllators of D=4 Yang-Mills theory can be derived from stochastic
(IF) quantization of the Chern-Simons action, as shown  in ref. \cite{HC},
and in detail in ref. \cite{BC}.

\section{Stabilized Gravity}

   To apply stochastic (IF) quantization to gravity, we must generalize
the formalism somewhat to allow for field-dependent  supermetrics.
Denote the fields, which may represent the metric, tetrad, or spin-connection,
by $g^N$, the supermetric by $G_{MN}$, and the supervielbien by
$E^A_N$.  The corresponding Langevin equation is \footnote{This form
corresponds to the $\m=\sigma=0$ case of the Langevin equation for gravity
in ref. \cite{Halpern-Chan}.  The most general case will not be considered
here.}

\beq
       \partial_5 g^M = - G^{MN} {\d S \over \d g^N} + E^M_A \eta^A
\eeq
where

\bea
          <\eta^A(\x,t_5) \eta^B(\x',t'_5)> &=& 2 \hbar \d_{AB}
\d^4(\x-\x') \d(t_5-t'_5)
\non \\
              G_{MN} &=& E_M^A E_N^A
\eea
The analogue of \rf{etaint} becomes

\bea
        Z_5 &=& \int Dg^N \; det[{\d \eta \over \d g}]
\; exp\left[-\int d^5x \; [ \oq G_{MN} \partial_5 g^M \partial_5 g^N
+ \oq G^{MN} {\d S \over \d g^M} {\d S \over \d g^N} ]/\hbar \right]
\non \\
& & \times e^{-\oh(S[g^N_f] -S[g^N_i])/2\hbar}
\label{analog}
\eea
and from the Langevin equation

\beq
        \eta^A = E^A_M (\partial_5 g^M + G^{MN} {\d S \over \d g^N})
\eeq
we have

\bea
  \lefteqn{   {\d \eta^A(x) \over \d g^L(x')}   }
\non \\
&=& E^A_M\left[\d^M_L \partial_5 \d^5(x-x') + {\d \over \d g^L(x')}
G^{MN}(x) {\d S \over \d g^N(\x)}_{|t_5}
- {\d E^A_M(x) \over \d g^L(x')} E^M_B \eta^B \right]
\non \\
           &=& E^A_M \partial_5 \left[ \d^M_L \d^5(x-x')
+ \int d\t \theta(t_5-\t) \{ {\d \over \d g^L(x')} G^{MN}(\x,\t)
{\d S \over \d g^N(\x)}_{|\t} \right.
\non \\
& & - \left. {\d E^M_B(\x,\t) \over \d g^L(x')}
\eta^B(\x,\t) \} \right]
\eea
leading to the determinant

\bea
    \lefteqn{ det[{\d \eta^A(x) \over \d g^L(x')}] }
\non \\
&=& det[\partial_5] det[E]
exp\left[\int d^5x d\t \theta(t_5-\t) [{\d \over \d g^M(\x,t_5)}
G^{MN}(\x,\t) {\d S \over \d g^N(\x)}_{|\t} \right.
\non \\
& &- \left. {\d E^M_B(\x,\t) \over \d g^M(\x,t_5)}  \eta^B(\x,\t) ] \right]
\label{det}
\eea

   The derivatives of the supermetric and supervielbein in \rf{det}
depend on a choice of stochastic calculus.  In the Ito calculus, all
contractions
between $G^{MN}(\x,t)$ and $\eta(\x',t)$ at equal (fifth) time $t$
are taken to be zero, e.g.

\bea
  \lefteqn{<E^N_A(\x,t) \eta^A(\x,t) E^M_B(\x',t) \eta^B(\x',t> }
\non \\
&=& <E^N_A(\x,t) E^M_B(\x',t)> <\eta^A(\x,t) \eta^B(\x',t)>
\eea
This condition can be achieved by defining

\beq
                     E^N_A(\x,t) \equiv E^N_A[g(\x,t-\d t)]
\eeq
where $\d t$ is an infinitesmal
fifth-time displacement, so that $E^N_A(\x,t)$ is independent
of $\eta(\x,t)$.  This fifth-time displacement does not affect invariance
under four-dimensional, $t_5$-independent, diffeomorphisms.
Extending this prescription also to the supermetric
in \rf{det}, i.e. $G^{MN}(\x,t)=E^M_A(\x,t) E^N_A(\x,t)$, results
in a determinant

\beq
    det[{\d \eta \over \d g}] =  det[\partial_5] det[E]
exp\left[\oh \int d^5x \; G^{MN}(x) {\d^2 S \over \d g^M(\x) \d g^N(\x)}_{|t_5}
\right]
\label{det3}
\eeq
Finally, substituting \rf{det3} into \rf{analog}, we obtain the
fifth-time action formulation

\bea
        <Q> &=& {1 \over Z_5}\int Dg^N \; det[E] Q[g^N(x,0)] e^{-S_5/\hbar}
\non \\
    S_5 &=& \int d^5x \; [ \oq G_{MN} \partial_5 g^M \partial_5 g^N
+ \oq G^{MN} {\d S \over \d g^M} {\d S \over \d g^N} - {\hbar \over 2}
G^{MN} {\d^2 S \over \d g^M \d g^N}]
\label{S5general}
\eea

   We now apply this formulation to the Einstein-Hilbert and
Einstein-Cartan actions.  The Einstein-Hilbert action is

\beq
       S_{EH} = - {1 \over \k^2} \int d^4x \sqrt{g} R
\label{SEH}
\eeq
where $\k^2 = 16 \pi G$.  Expanding $g_{\m \n} = \d_{\m\n} + \k h_{\m\n}$,
the action to zeroth-order is

\beq
       S^0 = \int \zp \; h_{\m \n}(p) p^2
[\oq P^{(2)} - \oh P^{(0-s)}]_{\m \n \a \b} h_{\a \b}(-p)
\eeq
where $P^{(0-s)}$ and $P^{(2)}$ are transverse spin-2 and spin-0
projection operators \cite{vanN}.  In this case
the 10 independent fields $g^A$ just correspond to the metric components
$g_{\m \n}$ ($\m \ge \n$).  The supermetric $G_{MN}$ is defined implicitly
from

\bea
      \d g^2 &=& \int d^4x  \; G_{MN}(x) \d g^M(x) \d g^N(x)
\non \\
          &=& \int d^4x \;  G^{\m \n \a \b}(x) \d g_{\m \n}(x)
\d g_{\a \b}(x)
\non \\
      G^{\m \n \a \b} &=&  \oh \sqrt{g} [g^{\m \a} g^{\n \b}
+ g^{\m \b} g^{\n \a} + c g^{\m \n} g^{\a \b} ]
\label{DeWitt}
\eea
It is required that the arbitrary constant $c$
in the DeWitt supermetric $G^{\m\n\a\b}$ be constrained to $c > -\oh$;
since otherwise $det(G)<0$ and we cannot construct a supervielbein.
This would break the link between stochastic (IF) quantization and
the 5-th time action (in fact, $S_5$ would no longer be bounded from
below).  Applying \rf{S5general} to the Einstein-Hilbert action
\rf{SEH}, one finds

\bea
     S_5 &=& \oq \int d^5x \; \left[ {1 \over \k^2}
G^{\m \n \a \b} \partial_5 g_{\m \n}
\partial_5 g_{\a \b} + \k^2  G_{\m \n \a \b}^{-1}
{\delta S \over \delta g_{\m \n}} {\delta S \over \delta g_{\a \b}} \right.
\nonumber\\
 & & \left. - 2 \hbar \k^2  G_{\m \n \a \b}^{-1} {\delta^2 S \over \delta
g_{\m \n}   \delta g_{\a \b}} \right]
\nonumber \\
         &=& \oq \int d^5x \; \left[ {1 \over \k^2}
G^{\m \n \a \b} \partial_5 g_{\m \n}
\partial_5 g_{\a \b} + {1 \over  \k^2} g G_{\m \n \a \b}^{-1}
(R^{\m\n}-\oh g^{\m\n} R) \right.
\nonumber \\
     & & \left. \times (R^{\a\b} - \oh g^{\a\b} R)
    - \hbar \b  \sqrt{g} R \right]
\label{S5EH}
\eea
for the corresponding 5-th time action, where $\beta$ is a singular
constant.  This action was derived
in \cite{Me1}, except that the operator ordering in the last term
of $S_5$, as well as the functional integration measure, was left
undetermined.  Stochastic (IF) quantization determines the integration
measure to be the (D=4 dimensional) DeWitt measure
$det(E) = \sqrt{G} = const$, while the "retarded supervielbien"
$E^A_M(x,t) \equiv E^A_M[\p(x,t-\d t)]$ suggested by the Ito calculus
determines the ordering of supermetric and functional derivatives shown
above.  This ordering will be particularly convenient for the Einstein-Cartan
theory.

   In ref. \cite{Me1} it was shown how to calculate the stabilized
4-dimensional action $S_{eff}$ perturbatively, starting from $S_5$.
At linearized level

\beq
       S_{eff}^0[g_{\m\n}] = \int \zp \; h_{\m \n}(p) p^2 \left[\oq P^{(2)} +
  \oh P^{(0-s)}\right]_{\m \n \a \b} h_{\a \b}(-p)
\label{Seff-g}
\eeq
which, like $S^0$ above, is transverse; only the sign of
the "wrong-sign" conformal mode has been flipped.  $S_{eff}$ will be non-local
at higher-orders, but it is guaranteed to have the same classical equations
of motion as $S_{EH}$.

     The problem with \rf{S5EH}, which is a shortcoming of the
5-th time approach in general, is that $S_5$ contains higher-derivative
terms, in this case proportional to $R^2$.  This means that reflection
positivity for reflections across the "ordinary" time ($x_4$) axis
cannot be guaranteed, which is problematic for continuing
$S_{eff}$ to Minkowski space.\footnote{Reflection positivity is not
ruled out either, since $S_5$ certainly has this property for bounded
$S$ despite the higher derivative terms.}
The presence of higher derivatives in
$S_5$ can be traced to the 2nd derivative terms in $S$.
For this reason, it was suggested in
\cite{Me1} to stabilize the Einstein-Cartan theory, which, like
the Chern-Simons theory discussed in the last section, contains only
first order derivatives.  The "fifth"-time action corresponding
to D=3 Chern-Simons theory is D=4 Yang-Mills theory, which is certainly
reflection positive across any axis.  As the Einstein-Cartan theory
has a tensor structure similar to that of Chern-Simons theory, there is
reason to expect that the stabilized version is also reflection positive.

   The Einstein-Cartan theory in 4 dimensions has the action
\beq
      S_{EC} = - {1 \over 4 \k^2} \int  \e_{abcd} e^a \land e^b \land
(d \w^{cd} +\w^{cf}\land \w^{fd})
\label{SEC}
\eeq
In order to write down a single Langevin equation for this system,
it is necessary to have the tetrad $e^a_\m$ and spin connection
$\w^{ab}_\m$ in the same multiplet $g^L$, and it is convenient to
rescale these fields so that they have the same dimensions.  The
dimensional quantities in the theory are $\k,\hbar$, so we rescale

\bea
     \et &=& {1 \over \k}e
\non \\
     \wt &=& \sqrt{\hbar} \w
\eea
with supermetric defined implicitly by
\footnote{This is not the most general possible $e-\w$ supermetric. Only
the simplest version of the stabilized theory will be considered here.}

\bea
     \d g^2 &=& \d \et^2 + \d \wt^2
\non \\
    &=& \int d^4x \; \sqrt{g} g^{\m\n} [ \d \et^a_\m \d \et^a_\n
+ \d \wt^{ab}_\m \d \wt^{ab}_\n]
\label{supermetric}
\eea
Applying this supermetric to the general formula \rf{S5general},
and noting that in this case (as in Chern-Simons theory)
the singular term
\beq
      G^{MN} {\d^2 S \over \d g^M \d g^N} = 0
\eeq
vanishes, the stabilized Einstein-Cartan theory (including a cosmological
term \linebreak $\int d^4x \l det(e)$) is found to be

\beq
     <Q[e,\w]> = {1 \over Z_5} \int De D\w \;
\left[ \prod_{x,t_5} det^{10}(e) \right]
\; Q[e(x,0),\w(x,0)] e^{-S_5/\hbar}
\eeq
where
\bea
   S_5 &=& \oq \int d^5x \sqrt{g} \left[
{1 \over \k^2} g^{\m\n} (\partial_5 e^a_\m \partial_5 e^a_\n + \hbar
\partial_5 \w^{ab}_\m \partial_5 \w^{ab}_\n) \right.
\non \\
& & + 4({1\over \k^2}R^{a}_{\m} R^{a}_{\n} g^{\m\n}
-\l R + \k^2 \l^2)
\non \\
 & & \left. + {1 \over  \k^4 \hbar} T^a_{\m \n} T^b_{\r \s}
(\d_{ab} g^{\m\r} + 2 e^\m_a e^\r_b)g^{\n\s} \right]
\label{S5EC}
\eea
is the 5-th time action and

\bea
       R &=& d\w + \w \wedge \w
\non \\
       T &=& de + \w \wedge e
\eea
are the curvature and torsion two-forms respectively.  The fifth-time
action \rf{S5EC} contains no higher-derivative terms and can be
shown, in the lattice version discussed in the next section, to be
reflection positive across the $x_4$ axis.

   The perturbative expansion of \rf{S5EC}, at $\l=0$, is an expansion
around the classical $R=T=0$ solution:

\bea
      e^a_\m &=& \d^a_m + \k b^a_\m
\non \\
      \w^{ab}_\m &=& \wb^{ab}_\m(e) + \k^2 \sqrt{\hbar} \Omega^{ab}_\m
\eea
where

\beq
      \wb^{ab}_\m = \oh e^{\n a}(\partial_\m e^b_\n - \partial_\n
e^b_\m ) - \oh e^{\n b}(\partial_\m e^a_\n - \partial_\n e^a_\m )
- \oh e^{\r a}e^{\s b}(\partial_\r e_{\s c} - \partial_\s e_{\r c})
e^c_\m
\eeq
is the zero-torsion spin-connection.  Then the part of $S_5$ which
is zeroth-order in $\hbar$ is

\bea
   S_5 &=& \oq \int d^5x \sqrt{g} \left[ {1 \over \k^2} g^{\m\n}
\partial_5 e^a_\m \partial_5 e^a_\n  + 4({1\over \k^2}{\overline R}^a_\m
{\overline R}^a_\s g^{\m\s} - \l {\overline R} + \k^2 \l^2 ) \right.
\non \\
& & \left. + (\Omega \wedge e)^a_{\m\n} (\Omega \wedge e)^b_{\r \s} g^{\n\s}
(\d_{ab} g^{\m\r} + 2 e^\m_a e^\r_b) + O(\sqrt{\hbar}) \right]
\label{S5EC0}
\eea
where ${\overline R} = d\wb + \wb \wedge \wb$.  From \rf{S5EC0}
we see that torsion propagates only at loop level.
The D=4 stabilized action
$S_{eff}$ was calculated at zeroth-order in $\k$ (for $\l=0$)
in ref. \cite{Me1}, and it was found that

\beq
       S_{eff}^0[e] = \int \zp \; h_{\m \n} p^2 \left[\oq P^{(2)} +
  \oh P^{(0-s)}\right]_{\m \n \a \b} h_{\a \b}
\eeq
where $h_{\m\n} = b_{\m\n}+b_{\n\m}$ is the symmetric part of the
tetrad.  This result is identical to the zeroth order $S^0_{eff}$
obtained for the stabilized Einstein-Hilbert action.

\section{Numerical Simulation}

   There is no known lattice action for general relativity which
is exactly invariant under a continuous symmetry group analogous to
diffeomorphisms.  The best one can do for the Einstein-Cartan theory is to
preserve the invariance under the local Lorentz group (O(4) in Euclidean
space), and hope that the diffeomorphism invariance of the continuum
action can somehow be
recovered at a fixed point. Here we follow Menotti and Pelissetto
\cite{Menotti} in introducing a hypercubic lattice with link variables

\bea
    U_\m(n) &=& exp[aP_b \tau^b_\m(n) + \oh aJ_{bc} \wt^{bc}_\m(n)]
\nonumber \\
              &=& exp[aP_b \et^{b}_{\m}(n)] \; exp[\oh a J_{bc}
\wt^{bc}_{\m}(n)]
\nonumber \\
              &=& exp[P_b e^{b}_{\m}(n)] \; exp[\oh J_{bc} \w^{bc}_{\m}(n)]
\label{link}
\eea

\noindent where $a$ is the lattice spacing, and from here on
we set $\hbar=1$. $P_a$ and $J_{ab}$ are the
generators of the Euclidean Poincare group in the four-dimensional
spinor representation

\bea
        J_{ab} &=& \oq [\gamma_a,\gamma_b]
\non \\
        P_a &=& \oh \gamma_a(1+\gamma_5)
\eea
Define the plaquette variable

\beq
       U_{\m\n}(n) = U_\m(n) U_\n(n+\m) U^{-1}_\m(n+\n) U^{-1}_\n(n)
\eeq
\noindent and lattice curvature, torsion, and tetrad

\bea
       R^{ab}_{\m\n}(n) &=& - \oh Tr\{J_{ab}[U_{\m\n}(n)-U_{\n\m}(n)] \}
\nonumber \\
       T^a_{\m\n}(n)    &=& - \oq Tr\{K_a [U_{\m\n}(n)-U_{\n\m}(n)] \}
\non \\
       e^a_\m &=& -\oq Tr\{K_a U_\m(n) \gamma_5 U^{-1}_{\m}(n) \}
\label{curvature}
\eea

\noindent where $K_a = - \oh \gamma_a(1-\gamma_5)$.
Under $O(4)$ gauge transformations, the lattice tetrad,
curvature, and torsion transform like vectors and tensors in the latin
(local frame) indices.

   In terms of these variables, the lattice version of the Einstein-Cartan
action becomes

\beq
       S = \sum_{n} \left[ - \oq \e^{\m\n\a\b} \e_{abcd} R^{ab}_{\m\n}(n)
e^{c}_{\a}(n) e^{d}_{\b}(n) + \l det(e) \right]
\eeq

\noindent which is invariant under local Lorentz O(4) gauge
transformations, but still bottomless.  The lattice cosmological
constant $\l$ is expressed in units of the Planck length, i.e.
$\l = \k^4 \l^{continuum}$; note that the lattice spacing has dropped
out of the action. In terms of the same variables,
the latticized version of the Einstein-Cartan 5-th time action
\rf{S5EC} is

\bea
     S_5 &=& \oq \sum_n  |det(e)| \left[ {1\over \e} e^\m_c(n) e^\n_c(n)
\{T_{\m5}^a(n) T_{\n5}^a(n) + R_{\m5}^{ab}(n) R_{\n5}^{ab}(n) \} \right.
\nonumber \\
        & & + 4 \e \{ R_{\m\n}^{ab} R_{\r\s}^{ac}
e^{\m}_d e^\r_d e^\n_b e^\s_c - \l R^{ab}_{\m \n}e^\m_a e^\n_b + \l^2 \}
\nonumber \\
        & & \left. + \e  T^a_{\m\n} T^b_{\r\s} (\delta_{ab}
e^\m_c e^\r_c + 2 e^\m_a e^\r_b) e^\n_d e^\s_d \right]
\label{S5lat}
\eea
where
\beq
      \e = {a_5 \over 2 \k^2}
\label{epsilon}
\eeq
and $a_5$ is the lattice spacing in $t_5$ direction.  The lattice
vectors $n$ now label sites on a D=5 dimensional
lattice, and $e^{\m}_a$ is
inverse to the matrix $e^a_\m$.  Defining $U_5(n)=1$, the curvature
and torsion $R_{\m5}^{ab}(n)$ and $T_{\m5}^a(n)$ are also obtained
from \rf{curvature}.  The quantities in \rf{curvature} are dimensionless;
the scalings which give the usual curvature, torsion, and tetrad in
the continuum limit are

\bea
      R^{ab(continuum)}_{\m\n} &=& {1 \over a^2} R^{ab}_{\m\n}
\non \\
      T^{ab(continuum)}_{\m\n} &=& {\k \over a^2} T^{ab}_{\m\n}
\non \\
      e^{a(continuum)}_\m &=& {\k \over a} e^a_\m
\eea

  The action \rf{S5lat}, together with the integration measure

\beq
       \int \prod_n det^{10}(e) \prod_\m dU_{\m}(n)
\eeq
was used for the numerical simulation.  Note that the lattice
spacing $a$ in the $1-4$ directions scales out of the action
completely, and all quantities which,
in the continuum, are diffeomorphism invariant, appear
on the lattice in units of $\k$ (c.f. \cite{mlen}), e.g.
\footnote{Disappearance of the lattice spacing $a$ in favor of
$\k$ is found also in the Regge lattice formulation.}
\bea
     \int d^4x \sqrt{g} R &=& \k^2 \sum_n |det(e)| R^{ab}_{\m\n}e^\m_a e^\n_b
\non \\
     \int d^4x \sqrt{g} &=& \k^4 \sum_n |det(e)|
\non \\
        \d s^2_{n,n+\d} &=& \k^2 <e^a_\m e^a_\n> \d^\m \d^\n
\eea
The expectation values of such quantities are therefore expressed, on
the lattice, in units of the Planck length $\k$ rather than lattice
spacing $a$, and depend only on
the parameters $\e = a_5/2\k^2$, and $\l$.  The fact that lattice
spacing $a$ drops out of both the action and the expectation values is
a remnant of diffeomorphism invariance in the continuum, and
has led to the speculation that lattice quantum gravity, like string
theory, has a kind of built-in minimum length $\d s^2 \approx \k^2$
\cite{mlen}.

   The final step in the lattice formulation above would be to
symmetrize \rf{S5lat} with respect to all $\pi/2$ rotations around
the axes; the resulting symmetrized action can be shown to be
reflection positive \cite{Menotti}.  In the interest of minimizing
computer time however, only the unsymmetrized action \rf{S5lat} was used
for the Monte Carlo simulation.

   Computer simulation of latticized Einstein-Cartan theory is rather
lengthy, even for very small lattices.  The link variables are $4 \times
4$ matrices on a 5-dimensional lattice,
and because of the plaquette-plaquette structure of the
action, a single link update requires on the order of $10^4$ floating-point
multiplications.  For this reason the simulation has only been carried
out on tiny lattices of dimensions $2^4 \times 4$ and $3^4 \times 4$
(4 spacings in the $t_5$ direction), which required, despite the
small lattice sizes, a total expenditure of 100 Cray YMP hours.

   There is a two-parameter ($\e - \l$) space of couplings for $S_5$.
One cannot vary $\e$ at fixed $\l=0$ because the lattice does not
seem to thermalize at $\l=0$, for reasons discussed below.  Instead
I have fixed $\e$ somewhat arbitrarily at $\e=\oh$, and carried out
Monte Carlo simulations for various values of $\l$.

   Figures 1-3 plot the average curvature (Fig. 1),

\beq
         <R> \equiv {<\sum |det(e)| R> \over <\sum |det(e)|>}
\eeq
the volume per lattice site (Fig. 2),

\beq
         <\sqrt{g}> \equiv {1 \over N_{sites}} <\sum_n |det(e)|>
\eeq
and the derivative of the free energy per lattice site (Fig. 3),

\beq
      {dF \over d\l} = {1 \over N_{sites}} <\sum_n |det(e)|
(-.5 R + \l)>
\eeq
versus the cosmological constant $\l$ in a range
$-5 \le \l \le 20$.  Squares denote data
on $2^4 \times 4$ lattices, at $\l=-1,.25,.5,1,2,5,10,20$,
crosses denote data on $3^4 \times 4$ lattices at
$\l=-5,-2,-1,.5,1,2,5$. The longest runs were at the smallest values of
$\l$, e.g. 300 thermalizations and 1600 data-taking iterations
at $\l=0.5$ on a $3^4\times 4$ lattice.

  From Fig. 1, we find that $<R>$ depends linearly
on $\l$, with $<R> \approx 1.6 \l$, as compared to the classical
value $R = 2 \l$.  The modest discrepancy in slope could have
various origins: renormalization of $\l$, the smallness of the
lattice, and the effect of the hypertoroidal lattice topology.

  Figure 2 shows a clear divergence in volume/site at $\l=0$.
Runs at the value $\l=0$ did not converge to a finite value
of $|det(e)|$, instead showing a steady increase in volume with number of
iterations.  Apart from the divergence at $\l=0$, there are
two other noteworthy facts concerning $<det(e)>$:  First, it is
found that for all values of $\l \le 10$, $<det(e)>=<|det(e)|>$,
while for the largest value of $\l=20$, $<det(e)>$ was only $4\%$
less than $<|det(e)|>$.  From this we conclude that in the range
of $\l$ studied, the system is in the broken symmetry phase expected
for quantum gravity, and it appears that fluctuations
which change the sign of $det(e)$  are rarely generated.\footnote{
Strictly speaking, transformations such as inversions $x_4 \ra -x_4$,
$U_4 \ra U^{-1}_4$, which take $det(e) \ra -det(e)$, are invariances of
the symmetrized, reflection positive lattice theory.  For the unsymmetrized
action of eq. \rf{S5lat}, used for the numerical simulation,
invariance under such transformations is only approximate.}
The second point is that the volume/site
decreases smoothly away from $\l=0$ for both positive and negative
$\l$.  This fact, and the fact that $<R> \propto \l$, show very clearly
the effect of stabilization, since the cosmological term in the
Einstein-Cartan/Hilbert actions is bottomless for $\l<0$.
In the fifth-time Einstein-Cartan,
changing the sign of the cosmological constant
(leaving Newton's constant fixed) is equivalent to changing the sign
of Newton's constant (leaving the cosmological constant fixed).  In the
Regge and simplicial manifold approaches, the behavior of average volume
and curvature changes drastically upon changing the sign of Newton's
constant [17-20].  This is not the case in the
stabilized version of the Einstein-Cartan theory,
where the metric fluctuates close
to zero curvature for small $\l$ of either sign.

   In connection with simplicial manifolds, it has been noted that the
lattice cutoff stabilizes the Euclidean Einstein-Hilbert action, since
the entropy of simplicial manifolds with large curvatures is small
\cite{Migdal}.  If continuum quantum gravity would then be stable due to
the measure, the 5th-time procedure cannot change anything, and is simply
a rewriting of the original 4-dimensional theory.  (This is because if
$exp(-S)$ is normalizable, then it is necessarily the ground state
(squared) of $H_5$; this means that $S_{eff}=S$, where $S$ is the
Einstein-Hilbert action.)  On the other hand, it may be that any
stability induced by the lattice cutoff in the D=4 theory
is spurious or non-universal,
in which case the 5th-time action is a much better starting point for
both numerical and analytical work.  Perturbative calculations in the
stabilized Einstein-Hilbert theory do indicate, in fact, that
$S_{eff} \ne S$, and indeed that $S_{eff}$ is non-local \cite{Me1},
which is in accord with the findings of ref. \cite{Schleich,Brout,Mazur}.

   Finally, we see in Fig. 3 that the slope of the free energy is large,
and possibly singular, in the neighborhood of $\l=0$.  Although the
data is nowhere near sufficient to allow a numerical integration,
it does appear that the negative free energy (and therefore $Z_5$)
must have a peak, and might even be divergent, at $\l=0$. This peak
is intriguing, because if the cosmological constant becomes dynamical
in some way, as suggested by Baum, Hawking and Coleman \cite{BHC},
then $Z_5(\l)$ has the interpetation of
a probability density for the cosmological constant, i.e.

\bea
       P(\l) &=& N \int Dg  \; e^{-S_{eff}[g,\l]}
\non \\
           &=& \int Dg \; \psi^*[g,\l] \psi[g,\l] \propto Z_5(\l)
\eea
A sharp peak in the (negative) free energy at $\l=0$ is therefore an
explanation
of the smallness of the cosmological constant.  It is too much to claim
that the data shown here, obtained on tiny lattices from an action
which breaks diffeomorphism invariance, is strong evidence for
the vanishing of the cosmological term in quantum gravity, but at least
the data does seem to support this idea.  Actually
the evidence for a sharp peak in $-F$ is better for $\l \ra 0^+$
than for $\l \ra 0^-$.  Conceivably, since the volume is divergent
at $\l=0$, $Z_5$ itself may be discontinuous at this point, with the
probability peaked at $\l=0^+$.  As usual, one would like to have
more data, on larger lattices, in the region of interest.

   I have made no attempt to search for an ultraviolet fixed point in the
$\e - \l$ parameter space; all computations at various $\l$
were made at the constant (and rather arbitrary) value of $\e =\oh$.
A fixed point where diffeomorphism invariance is restored (if such
a point exists) would be signaled by a peak, growing sharper with lattice
size, in the correllation

\beq
     C = {<(\sum \sqrt{g} R)^2> - <\sum \sqrt{g} R>^2
\over <\sum \sqrt{g}> }
\eeq
No such peak was observed for $\e = \oh$ on the $3^4 \times 4$
lattice; the value of $C \approx 11.5$ was almost constant, varying
by less than $10 \%$, over the full range of $\l$.  A very interesting
question is whether a peak in correllation $C$
develops as $\e$ is varied; this is an
issue where much additional numerical work is required.

\vspace{33pt}

\noindent {\Large \bf Acknowledgements}{\vspace{11pt}}

  I would like to thank Marty Halpern for some stimulating
discussions, and also to thank the LBL theory group for their
hospitality.   The computer calculations were supported by the
Florida State University
Supercomputer Computations Research Institute, which is partially funded
by the U.S. Department of Energy through Contract No.
DE-FC05-85ER250000.

\newpage

\noindent {\Large \bf Figure Captions}
\bigskip
\bigskip

\begin{description}
\item[Fig. 1] Average curvature vs. cosmological constant, in units
of the Planck length. Squares denote data on $2^4 \times 4$ lattices,
crosses denote data on $3^4 \times 4$ lattices.
\item[Fig. 2] $<\sqrt{g}>$ (volume/site) vs. cosmological constant.
\item[Fig. 3] Derivative of the free energy/lattice site
$dF/d\l$ vs. cosmological constant.
\end{description}

\end{document}